\documentclass[aps,twocolumn,longbibliography]{revtex4}
\usepackage{graphicx}
\begin{document}

\title{Topological phase transitions in tilted optical lattices}

\author{Andrey~R.~Kolovsky$^{1,2}$}
\affiliation{$^1$Kirensky Institute of Physics, 660036 Krasnoyarsk, Russia}
\affiliation{$^2$Siberian Federal University, 660041 Krasnoyarsk, Russia}
\date{\today}
\begin{abstract}
We analyze the energy spectrum and eigenstates of cold atoms in a tilted brick-wall optical lattice.  When the tilt is applied, the system exhibits a sequence of topological phase transitions reflected in an abrupt change of the eigenstates. It is demonstrated that these topological phase transitions can be easily detected in a laboratory experiment by observing Bloch oscillations of cold atoms. 
\end{abstract}
\maketitle

{\em 1.}  Recently much attention has been paid to topological properties of matter \cite{book}, where cold atoms in optical lattices provide an excellent playground for studying different models. For example, in  Ref.~\cite{Atal13} the authors realized the one-dimensional Su-Schrieffer-Heeger (SSH) model  \cite{Su79} and measured the Zak phases for two different dimerization of  the SSH lattice. In Ref.~\cite{Aide15} the two-dimensional lattice with effective magnetic field was created and the Chern number of the ground magnetic band was measured by observing the anomalous velocity of atoms. The cold atom realization of the Haldane lattice  \cite{Hald88} was reported in Ref.~\cite{Jotz14}.   

\begin{figure}
\includegraphics[width=7.0cm]{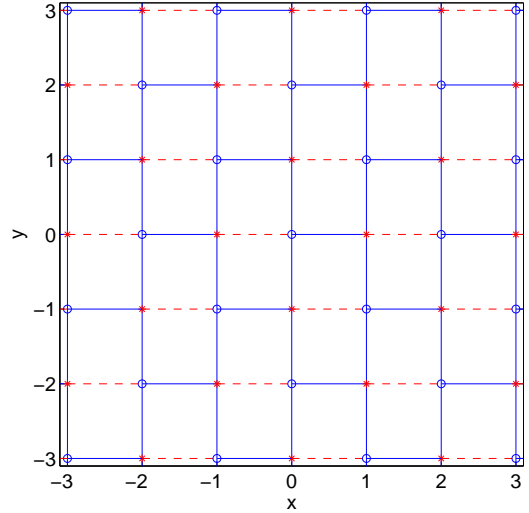}
\caption{The brick-wall lattice. The bonds marked by the solid and dashed lines corresponds to the different hoping matrix elements $J_1$ and $J_2$. Notice that for $J_2=0$ the brick-wall lattice transforms into the deformed honey-comb lattice with $A$ and $B$ sites marked by asterisks  and open circles.}
\label{fig0}
\end{figure}

In this work we discuss a new class of topologically nontrivial systems which should be treated as quasi one-dimensional. These are the tilted  bipartite lattices. For the first time topological phase transitions in tilted bipartite lattices were mentioned in Ref.~\cite{109} which studied the biased dice lattice. The dice lattice has a rather particular geometry that leads to the Bloch spectrum consisting of three bands, with one band being completely flat. While the emphasis of Ref.~\cite{109} was on flat bands, in the present work we focus on the topological phase transitions and their physical manifestations that can be detected in the present-day laboratory experiments with cold atoms. Putting laboratory accessibility on the first place, we consider the brick-wall lattice which was used earlier in the experiment \cite{Tarr12} to study the effect of Dirac cones. The network connectivity of the brick-wall lattice is shown in Fig.~\ref{fig0} where the solid and dashed bonds correspond to the strong $J_1$ and weak $J_2$ coupling, respectively. Here we analyze the simplest case where the brick-wall lattice is tilted in the $y$ direction. In other words, the atoms are subject to a potential field ${\bf F}$ aligned with the $y$ axis. The field  breaks translational invariance of the system in the $y$ direction and causes the atoms to be localized within the Stark localization length $\sim 1/F$. However, in the $x$ direction translational symmetry is preserved and, hence, atomic eigenstates are extended Bloch-like states. Localization in one direction together with translational invariance in the other relates the  system to the SSH model. However, due to the presence of an extra parameter $F$, the system is essentially richer than the celebrated SSH lattice. In particular, the tilted brick-wall lattice preserves its topological properties even if the hopping matrix element $J_2$ is strictly zero.

\begin{figure}
\includegraphics[width=9.5cm]{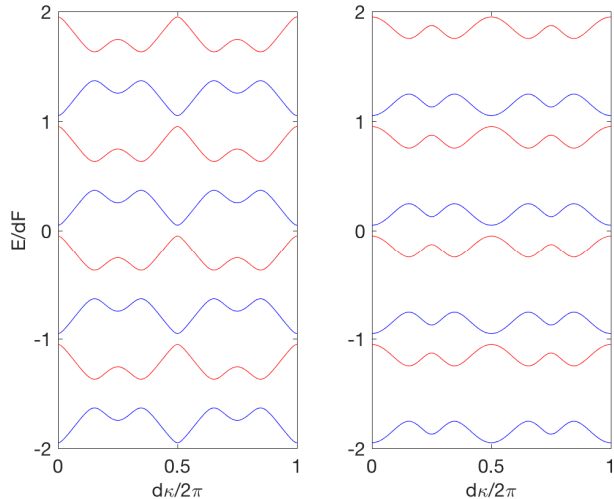}
\caption{Energy spectrum of the tilted brick-wall lattice. Parameters are $d=1$, $J_1=0.5$, $J_2=0$, and $F=1/1.9$ (left panel) and $F=1/2.1$ (right panel), which are slightly above and below $F_{cr}=1/2$ (see text for details).}
\label{fig1}
\end{figure}

{\em 2.} We begin with the energy spectrum of a quantum particle in the tilted brick-wall lattice. As for any bipartite lattice, it consists of pairs of one-dimensional  energy bands arranged into the Wannier-Stark ladder  \cite{101}, see Fig.~\ref{fig1}. Numerically this spectrum can be found by using at least two different methods  -- by diagonalizing the truncated Hamiltonian matrix,  or by calculating and diagonalizing  the Floquet operator, which is the system evolution operator over the Bloch period. The latter approach has certain advantages over the former because it (i) does not require truncation of the Hilbert space and (ii) opens prospects for analytical studies of the spectrum \cite{101,100}. 

\begin{figure}[b]
\includegraphics[width=8.0cm]{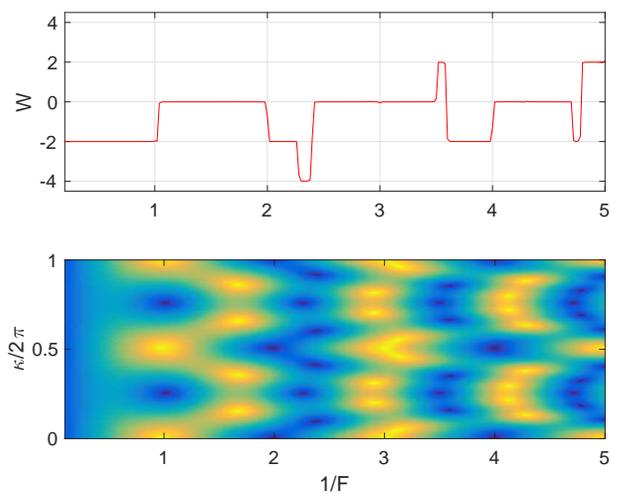}
\caption{Upper panel: Winding number of the phase $\chi$ vs. the inverse field magnitude. Lower panel: The energy band $E^{(+)}_{n=0}$ as a color map. The limits of the energy axis are $E_{min}=0$ (dark blue) and $E_{max}=dF/2$ (bright yellow).}
\label{fig3}
\end{figure}

In the framework of the Floquet operator the Wannier-Stark spectrum is given by the equation 
\begin{equation}
\label{b0}
E_n^{(\pm)}(\kappa)=dFn+i\frac{dF}{2\pi}\ln \lambda_{1,2}(\kappa)  \;,
\end{equation}
where $d$ is the lattice period and $\lambda_1$ and $\lambda_2=\lambda_1^*$ are eigenvalues of  the $2\times2$ unitary matrix 
\begin{equation}
\label{b1}
U(\kappa)=\widehat{\exp}\left[\frac{i}{dF}\int_0^{2\pi} G(\theta;\kappa){\rm d}\theta \right]  \;.
\end{equation}
The explicit form of the matrix $G(\theta;\kappa)$ in Eq.~(\ref{b1}) depends on the field orientation and in our case is given by
\begin{equation}
\label{b2}
G(\theta;\kappa)=\left(
\begin{array}{cc}
0 & f  \\
f^* & 0 
\end{array} \right) \;,
\end{equation}
where $f(\theta;\kappa)=J_2+J_1 (e^{i2d\kappa}+2e^{id\kappa}\cos\theta)$.  As follows from Eq.~(\ref{b1}) the Wannier-Stark energy bands are sensitive to variation of $F$.  For future reference the lower panel in Fig.~\ref{fig3} shows the blue band with the ladder index $n=0$  as a function of the quasimomentum and inverse field magnitude. Our particular interest in this figure are points of conical intersections between energy bands of different symmetry, seen as the dark and bright spots.

{\em 3.} We proceed with topological phase transitions. The  main object to study are eigenvectors ${\bf Y}_{1,2}(\kappa)=[Y^A_{1,2}(\kappa), Y^B_{1,2}(\kappa)]^T$   of the unitary matrix (\ref{b1}), whose component determine occupation probabilities of the $A$ and $B$ sites of the brick-wall lattice. Due to particular algebraic structure of the matrix (\ref{b2}), the eigenvectors have the following simple form:
\begin{equation}
\label{b3}
{\bf Y}_{1,2}=\left(
\begin{array}{c}
\frac{1}{\sqrt{2}}e^{i\chi} \\  \pm \frac{1}{\sqrt{2}} e^{-i\chi}
\end{array}
\right) \;,
\end{equation}
where $\chi$ is a function of the quasimomentum and without any loss of generality we can require $\chi(\kappa=0)=0$. We also take a convention that index `1' always refer to the upper (blue) band and index `2' to the lower (red) band. Following Ref.~\cite{109} we introduce a topological invariant -- the winding number $W$ \cite{remark0},
\begin{equation}
\label{b4}
W=\frac{1}{2\pi i} \int_0^{2\pi} {\rm angle}\left(\frac{Y^A_j}{Y^B_j}\right) {\rm d}\kappa
=\frac{1}{\pi i} \int_0^{2\pi} e^{-i\chi}\frac{{\rm d}}{{\rm d} \kappa} e^{i\chi} {\rm d}\kappa \;,
\end{equation}
which is an integer number because ${\bf Y}_{1,2}(\kappa)$ are periodic functions of the quasimomentum. Notice that $W$ is not affected by gauge transformations when the eigenvectors are multiplied by a phase factor $\exp[i \Phi(\kappa)]$ with $\Phi(\kappa)$ being an arbitrary periodic function of $\kappa$. As $F$ is varied the winding number (\ref{b4}) may change its value only at some critical values of the field, where energy bands of different symmetry touch each other, see upper panel in Fig.~\ref{fig3}. The  observed jumps of the winding number is a formal manifestation of topological phase transitions.  

Let us focus on the single topological transition, say at $F_{cr}=1/2$, where the bands shown in Fig.~\ref{fig1} touch each other at $\kappa=0$ and $\kappa=\pi$. In vicinity of this critical value the matrix (\ref{b1}) is close to the identity matrix,
\begin{equation}
\label{b5}
U(\kappa=0)=\left(
\begin{array}{cc}
\sqrt{1-\epsilon^2} & i\epsilon  \\
i\epsilon  & \sqrt{1-\epsilon^2}
\end{array} \right) \;,
\end{equation}
where $\epsilon$ is proportional to $\Delta F=F-F_{cr}$. It is easy to see that for $\epsilon \ll 1$ eigenvectors of the matrix (\ref{b5}) are given by Eq.~(\ref{b3}) with $\chi=0$ and eigenvalues by $\lambda_{1,2}\approx \exp(\pm i \epsilon)$.  The crucial point is that off-diagonal elements of the matrix (\ref{b5}) change their sign when $F$ crosses $F_{cr}$. Then the symmetric eigenvector ${\bf Y}_1=[1/\sqrt{2}, 1/\sqrt{2}]^T$, which has been associated with the upper band, skips to the lower band \cite{remark}.  This seemingly simple result has important consequences which can be detected experimentally by inducing Bloch oscillations of atoms in the $x$ direction. 
\begin{figure}
\includegraphics[width=8.0cm]{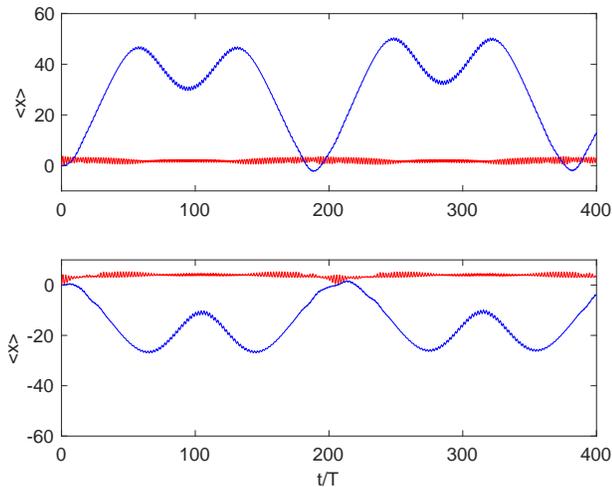}
\caption{Bloch oscillations of a localized wave packet for $F=1/1.9$ (upper panel) and $F=1/2.1$ (lower panel). Time is measured in units of the tunneling period. The mean value $x(t)=\langle \psi(t)| \hat{x} | \psi(t)\rangle$, blue curves,  and $y(t)=\langle \psi(t)| \hat{y} | \psi(t)\rangle$, red curves. The ratio of the field components $F_x/F_y=0.005$. } 
\label{fig5}
\end{figure}

{\em 4.}  We  consider a BEC of non-interacting cold atoms (i.e., a coherent wave-packet with vanishing phase difference between different sites)  and apply an external field ${\bf F}$ which is slightly mismatched with the $y$ axis, $F_x \ll F_y\approx F$. In this setup the strong component $F_y$ creates  the ladder of Wannier-Stark energy bands and the weak component $F_x$ induces Bloch oscillations of atoms in these bands. These oscillations can be conveniently described in terms of the Wannier-Stark states   $|\Psi_n^{(\pm)}(\kappa)\rangle$ as soon as we specify  the expansion coefficient $a_n^{(\pm)}(\kappa)=\langle \Psi_n^{(\pm)}(\kappa) | \psi(t=0) \rangle$. Since the occupation amplitudes of $A$ and $B$ sites in the initial state $|\psi(t=0)\rangle$ have equal phases, we predominantly populate $\kappa=0$ vicinity of blue bands if $F_y<F_{cr}$,  and  $\kappa=0$ vicinity of red bands if $F_y>F_{cr}$. Thus, the weak component $F_x$ induces Bloch oscillations either in red or blue bands depending on inequality relation between $F$ and $F_{cr}$. This is illustrated in Fig.~\ref{fig5} which shows the results of straightforward numerical simulations of the wave-packet dynamics for $F_x/F_y=0.005$ and $F=1/1.9$, upper panel, and $F=1/2.1$, lower panel.  It is seen that  the wave-packet displacement in the $x$ direction reproduces the dispersion relation of blue bands in Fig.~\ref{fig1}(a)  and red bands in Fig.~\ref{fig1}(b), respectively. (Since $dF_y>J_1$ the wave-packet displacement in the $y$ direction can be neglected.) In our simulations we intentionally use very small ratio $F_x/F_y$ since for a larger ratio the Bloch oscillations become complicated by interband Landau-Zenner transitions. Yet, qualitatively the effect remains the same -- the wave-packet moves in opposite directions for $F$ smaller and larger than $F_{cr}$.

{\em 6.}  In summary, we analyzed topological phase transitions in the tilted brick-wall optical lattice which are induced by variation of the field magnitude. These transitions correspond to a qualitative change of the two-dimensional Wannier-Stark states and can be detected by observing atomic Bloch oscillations, which nowadays is a routine procedure in the cold-atom physics. We notice, in passing, that one gets similar results for the more familiar honeycomb lattice.

To conclude the work we mention that the discussed phase transitions can be also observed in the curved two-dimensional photonic crystals, where the inverse curvature radius plays the role of the field  magnitude \cite{Trom06,Szam07,Corr13}. The latter system also admits to study the other manifestation of topological phase transitions, namely, a rearrangement  of the edge states. The detailed analysis of this rearrangement will be given in a separate paper.

{\em Acknowledgements.} The authors acknowledge fruitful discussions S.~Flach and and D.~N.~Maksimov and financial support from Russian Foundation for Basic Research, Government of Krasnoyarsk Territory, and Krasnoyarsk Region Science and Technology Support Fund through the joint grant No. 16-42-240746.



\begin{thebibliography}{10}
\bibitem{remark0}
Introduced winding number should not be mixed with that discussed in Ref.~\cite{Lee15}. The former is a characteristic 
of the eigenstates of tilted bipartite lattices while the latter is a characteristic of the spectrum of tilted topological lattices (for example, the Haldane lattice) in the limit $F\rightarrow 0$.
\bibitem{remark}
We mention that this result, when taken together with the fact that eigenvectors are periodic functions of the quasimomentum, necessary  implies a change of the winding number. The latter happens through developing of a non-analyticity of $\chi=\chi(\kappa)$ at $\kappa=0$ and $\kappa=\pi$.
\end{thebibliography}
\end{document}